\documentclass[nofootinbib,notitlepage,onecolumn,secnumarabic,amssymb, nobibnotes, superscriptaddress,aps, prd]{revtex4-1}
\usepackage{graphicx,epsfig}
\pdfoutput=1 
\usepackage{upgreek}
\usepackage{amsmath}
\usepackage{amssymb}
\usepackage{cancel}
\usepackage{amsfonts}
\usepackage{xcolor}
\usepackage{hyperref}
\usepackage{bm}
\usepackage{epstopdf}
\usepackage{natbib}
\usepackage{hyperref}

\newcommand{\cH}{\mathcal{H}}

\newcommand{\ud}{{\mathrm{d}}}

\def\k{\bm{k}}
\def\be{\begin{equation}}
\def\ee{\end{equation}}
\def\bea{\begin{eqnarray}}
\def\eea{\end{eqnarray}}

\newcommand{\bsb}{\boldsymbol}

\def\k{{\bsb k}}

\def\cH{\mathcal{H}}

\def\bk{{\bsb k}}

\usepackage{colortbl}
\definecolor{summersky}{cmyk}{0.71,0.33,0,0.14}
\definecolor{flamingo}{cmyk}{0,0.51,0.71,0.14}
\definecolor{rp}{cmyk}{0.2, 1, 0.6, 0}
\definecolor{pacificblue}{cmyk}{0.95,0.3,0, 0.19}
\definecolor{gray60}{cmyk}{0.4,0.4,0,0.8}
\definecolor{green94}{cmyk}{.94,0,1.00,0}
\definecolor{green80}{cmyk}{.80,0,.90,0}
\DeclareMathSizes{10}{10}{9}{9}
\title{Galaxy bispectrum with vectors and tensors}

\begin{document}
\title{\large Imprints of local lightcone projection effects on the galaxy bispectrum\\ {IV}: Second-order vector and tensor contributions }
\date{\today}

\author{Sheean Jolicoeur}
\email{beautifulheart369@gmail.com}
\affiliation{Department of Physics \& Astronomy, University of the Western Cape, Cape Town 7535, South Africa}

\author{Alireza Allahyari}
\email{allahyari@physics.sharif.edu}
\affiliation{Department of Physics, Sharif University of Technology, Tehran, Iran, and
School of Astronomy, Institute for Research in Fundamental
Sciences (IPM), P. O. Box 19395-5531, Tehran, Iran}

\author{Chris Clarkson}
\email{chris.clarkson@qmul.ac.uk}
\affiliation{School of Physics \& Astronomy, Queen Mary University of London, UK}
\affiliation{Department of Physics \& Astronomy, University of the Western Cape, Cape Town 7535, South Africa}
\affiliation{Department of Mathematics and Applied Mathematics,
University of Cape Town,
Rondebosch 7701, South Africa}

\author{Julien Larena}
\email{julien.larena@uct.ac.za}
\affiliation{Department of Mathematics and Applied Mathematics,
University of Cape Town,
Rondebosch 7701, South Africa}

\author{Obinna Umeh}
\email{umeobinna@gmail.com}
\affiliation{Institute of Cosmology \& Gravitation, University of Portsmouth, Portsmouth PO1 3FX, United Kingdom}
\affiliation{Department of Physics \& Astronomy, University of the Western Cape, Cape Town 7535, South Africa}

\author{Roy Maartens}
\email{roy.maartens@gmail.com}
\affiliation{Department of Physics \& Astronomy, University of the Western Cape, Cape Town 7535, South Africa}
\affiliation{Institute of Cosmology \& Gravitation, University of Portsmouth, Portsmouth PO1 3FX, United Kingdom}

\begin{abstract}

The galaxy bispectrum on  scales around and above the equality scale receives contributions from relativistic effects. Some of these arise from lightcone deformation effects, which come from local and line-of-sight integrated contributions. Here we calculate the local contributions from the generated vector and tensor background which is formed as scalar modes couple and enter the horizon. We show that these modes are sub-dominant when compared with other relativistic contributions.

\end{abstract}

\maketitle

\section{Introduction}

 The galaxy bispectrum will help differentiate between cosmological models in future surveys \cite{Sefusatti:2006pa,Sefusatti:2007ih,Komatsu:2010hc,Liguori:2010hx,Yadav:2010fz,Baldauf:2010vn,Yokoyama:2013mta,Camera:2014bwa,Abramo:2017xnp,Karagiannis:2018jdt}, test the equivalence principle on large scales \cite{Creminelli:2013nua}, further constrains the modified gravity models \cite{Hashimoto:2017klo} and will help constrain primordial non-Gaussianity \cite{Tellarini:2015faa, Tellarini:2016sgp, DiDio:2016gpd}.
 On large scales the bispectrum is affected by relativistic effects, mainly from lightcone projection effects on the observed number density~\citep{Challinor:2011bk, Bonvin:2011bg, Bonvin:2014owa,DiDio:2015bua, Kehagias:2015tda,Bertacca:2017dzm,Koyama:2018ttg,Umeh:2016nuh,Jolicoeur:2017nyt,Jolicoeur:2017eyi}. These effects can mimic some aspects of primordial non-Gaussianity, and need to be fully understood to extract the primordial signal from further surveys. This paper forms part of a series investigating these effects~\cite{Umeh:2016nuh,Jolicoeur:2017nyt,Jolicoeur:2017eyi}, which is complimentary to other approaches~\cite{DiDio:2016gpd,Bertacca:2017dzm}.

In this work we extend the previous studies of the bispectrum to include local vector and tensor perturbations. These vector and tensor perturbations are inevitably generated by scalar modes at second order \citep{Tomita:2005et,Mollerach:2003nq, Ananda:2006af, Lu:2007cj, Baumann:2007zm, Lu:2008ju, Hwang:2017oxa}. In fact the change to the metric from these contributions can reach nearly 1\% in the case of vectors~\cite{Lu:2007cj}, which has been confirmed in numerical simulations~\cite{Bruni:2013mua, Adamek:2015eda}. Frame dragging, like tensor perturbations,  doesn't couple strongly to matter, but it can change gravitational lensing~\cite{Andrianomena:2014sya, Saga:2015apa} and other lightcone distortion effects in the galaxy correlation function~\cite{Durrer:2016jzq, Bonvin:2017req, Tansella:2018hdm}, for example. 
Our analysis, while focused on these secondary modes, can also be adapted to other sources of vectors and tensors. 

\section{Galaxy bispectrum with local vectors and tensors}

Assuming a flat Friedmann-Lema\^itre-Robertson-Walker (FLRW) background, we write the metric in the Poisson gauge and conformal time $\eta$ as 
\begin{align}
\ud s^2=a^2\bigg[ -\Big(1+2\Phi^{(1)}\Big)\ud\eta^2+\omega_{i}^{(2)}\ud x^{i}\ud\eta+\Big[\Big(1-2\Psi^{(1)}\Big)\delta_{ij}+\frac{1}{2}h_{ij}^{(2)}\Big]\ud x^{i} \ud x^{j}\bigg]\;, \label{eq:2} 
\end{align}
where we only keep first order scalar perturbations sourcing the second order vector and tensor perturbations, $\omega^{(2)}_{i}$ and $h_{ij}^{(2)}$ respectively; the first order vector and tensor modes, which are purely decaying modes, are neglected \citep{Saga:2015apa, Bojowald:2007hv, Dai:2013kra}. Hereafter we drop the (2) superscript on $\omega_i$ and $h_{ij}$, and we also have $\Phi^{(1)}=\Psi^{(1)}=\Phi$. Vector and tensor modes are characterised by being divergence-free.

The solution for $\omega_{i}(\bm{x})$ is given by \citep{Lu:2008ju},
\begin{equation}
\nabla^{2}\omega_{i}(\bm{x}) = \frac{16f}{3\Omega_{m}\cH}\Big[\nabla^{2}\Phi(\bm{x})\partial_{i}\Phi(\bm{x})\Big]^{\mathrm{V}}\;, \label{eq:wi}
\end{equation}
where V denotes the vector part (i.e, the part of a 3-vector which cannot be written as the gradient of a scalar). Here, $f = \ud\ln D/\ud\ln a$ is the derivative of the linear growth rate, and $\Omega_m=\Omega_{m0}H_0^2/(a\cH^2)$.

The other contribution at second-order is the generation of tensor modes as structure forms, considered in~\cite{Matarrese:1993zf,Mollerach:2003nq, Noh:2004bc,Tomita:2005et,Ananda:2006af,Baumann:2007zm,Malik:2008im,Hwang:2012aa,Rampf:2013dxa,Andrianomena:2014sya,Osano:2015pea,Rampf:2014mga,Saga:2015apa}. These are found from projecting out the tensor part of the $ij$ part of the EFE. This gives (adapted from~\cite{Baumann:2007zm} to include a cosmological constant)
\bea
h_{ij}'' + 2{\cal H} h_{ij}'
-\nabla^{2} h_{ij}  = S_{ij}(\bm x)&=&
8
\bigg[\partial_i\Phi\partial_j\Phi
+\frac{1}{4\pi G a^2\rho}\partial_i(\Phi'+\cH\Phi)\partial_j(\Phi'+\cH\Phi)
\bigg]^\mathrm{T}\,,
\\
&=&8\left[1+\frac{2f^2}{3\Omega_m}\right]\big[\partial_i\Phi\partial_j\Phi\big]^\mathrm{T}\,,
\label{sdjncsdjkcn}
\eea
where $T$ denotes a  tensor projection. Note we have used $\big[\Phi\partial_i\partial_j\Phi\big]^\mathrm{T} = -\big[\partial_i\Phi\partial_j\Phi\big]^\mathrm{T}$.

\subsection{Second order galaxy number density}

We expand the observed galaxy number density to second order as $\Delta_{g}=\Delta_{g}^{(1)}+\frac{1}{2}\Delta_{g}^{(2)}$.
The expression for $\Delta_{g}$ comprising all the relativistic terms to second order is derived in  Ref.~\citep{Bertacca:2014dra,Bertacca:2014wga,Bertacca:2014hwa,Nielsen:2016ldx}. Further, the second order galaxy number density is expanded in terms of Newtonian and relativistic parts as 
\begin{equation}
\Delta_{g}^{(2)}=\Delta_{g\mathrm{N}}^{(2)}+\Delta_{g\mathrm{S}}^{(2)}+\Delta_{g\mathrm{V}}^{(2)}+\Delta_{g\mathrm{T}}^{(2)}\;, \label{eq:1}
\end{equation}
where $\Delta_{g\mathrm{V}}^{(2)}$ and $\Delta_{g\mathrm{T}}^{(2)}$ are the contribution to the number count from second order (generated by scalars) vector and tensor perturbations respectively. The expression for the Newtonian part, $\Delta_{g\mathrm{N}}^{(2)}$, can be found in \citep{Umeh:2016nuh}, and the local parts of the second-order non-Newtonian scalar contribution, $\Delta_{g\mathrm{S}}^{(2)}$, have been investigated in~\cite{Jolicoeur:2017nyt,Jolicoeur:2017eyi}. 

One can show that the second-order local contribution of vectors to the galaxy number count is given by \citep{Bertacca:2014hwa}:
{
\begin{align}
\Delta_{g\mathrm{V}}^{(2)}(z, \bm{x})=\left[-b_{e}+2\mathcal{Q} + \frac{2(1-\mathcal{Q})}{\chi\cH} + \frac{\cH'}{\cH^{2}}+\frac{1}{\mathcal{H}}\partial_{\|} \right] \omega_{\|}(\bm{x})\;.
\label{eq:3}
\end{align}
}
In \eqref{eq:3}, the radial comoving distance is denoted by $\chi=\eta_{0}-\eta$. The evolution bias is $b_{e}=\ud \ln (a ^3 n_{g})/\ud\ln{a}$ where $n_{g}$ is the galaxy number over-density, and $\cal Q$ is the magnification bias. The direction of observation is $\bm{n}$ and $\omega_{\|}=n^{i}\omega_{i}^{(2)}$ is the longitudinal component of the vector perturbation. Throughout the paper, a prime denotes a (partial) derivative with respect to $\eta$. Similarly, for the local tensor contribution, one finds \citep{Bertacca:2014hwa}: 
\begin{align}
 \Delta^{(2)}_{g\mathrm{T}}\left(\bm{x}\right)=-\frac{1}{2}(1-\mathcal{Q})h_\parallel\left(\bm{x}\right)-\frac{1}{2\cH}h'_{\parallel}\left(\bm{x}\right)\;, \label{eq:4}
 \end{align}
with $h_{\parallel}=h_{ij}^{(2)}n^{i}n^{j}$. In Fourier space, we may write \eqref{eq:3} and \eqref{eq:4} in terms of integrals over kernels which we derive in Appendix~\ref{AppA}:
\begin{eqnarray}
\Delta_{g\mathrm{V}}^{(2)}(\bk_{3}) &=& \int \ud(\bm{k}_{1}, \bm{k}_{2}, \bm{k}_{3}) \, \mathcal{K}_{\mathrm{V}}^{(2)} (\bk_{1},\bk_{2},\bk_{3})\;, \label{eq:11} \\
\Delta^{(2)}_{g\mathrm{T}}(\bk_{3}) &=& \int \ud(\bm{k}_{1}, \bm{k}_{2}, \bm{k}_{3}) \, \mathcal{K}_{\mathrm{T}}^{(2)} (\bk_{1},\bk_{2},\bk_{3})\;, \label{eq:13}
\end{eqnarray}
where we define the operator,
\begin{equation}
\int \ud(\bm{k}_{1}, \bm{k}_{2}, \bm{k}_{3}) = \int \frac{\ud^{3}k_{1}}{(2\pi)^{3}}\int \ud^{3}k_{2}\,\delta^{(1)}(\bm{k}_{1})\delta^{(1)}(\bm{k}_{2})\delta^{\mathrm{D}}(\bm{k}_{1}+\bm{k}_{2}-\bm{k}_{3})\;.\label{eq:12}
\end{equation}
Here, $\delta^{(1)}(\eta,\bk)$ is the first-order dark matter overdensity defined in the total-matter (T) gauge and $\delta^{\mathrm{D}}$ is the 3-dimensional Dirac-delta function. The explicit forms for the kernels in \eqref{eq:11} and \eqref{eq:13} are derived in Appendix~\ref{AppA}, and all geometric quantities associated with the coordinate system we use are defined in Appendix~\ref{appB}. For vectors we have 
\begin{eqnarray}
\mathcal{K}^{(2)}_{\mathrm{V}}(\bm{k}_{1}, \bm{k}_{2}, \bm{k}_{3} ) &=& 
12\Omega_{m}\cH^{2}f\bigg(\frac{\mu_3}{k_3}-\mathrm{i}\frac{\cH}{k_3^2}\bigg[-b_{e}+2\mathcal{Q}+\frac{2(1-\mathcal{Q})}{\chi\cH}+\frac{\cH'}{\cH^{2}}\bigg]\bigg)\nonumber\\&&
\times
\left[
\bm n\cdot \bm e(\bk_3)\left(\frac{\bk_1\cdot\bm e(\bk_3)}{k_1^2}+\frac{\bk_2\cdot\bm e(\bk_3)}{k_2^2}
\right)
+\bm n\cdot \bar{\bm e}(\bk_3)\left(\frac{\bk_1\cdot\bar{\bm e}(\bk_3)}{k_1^2}+\frac{\bk_2\cdot\bar{\bm e}(\bk_3)}{k_2^2}
\right)
\right]\,,
\eea
where $\bm e$ and $\bar{\bm e}$ are the 2 polarisation vectors associated with the vector mode.   
The kernel is complex and symmetric on $\bk_1\leftrightarrow\bk_2$. The $k$-dependence of the real part is $\mathcal{O}(k^{-2})$ and that of the imaginary part is $\mathcal{O}(k^{-3})$. We can see that for the equilateral configuration i.e., $k_{1} = k_{2}=k_{3} = k$, this kernel is zero, on using $\bk_1\cdot\bm e(\bk_3) = -\bk_2\cdot\bm e(\bk_3)$. Physically this is because the bispectrum in the equilateral configuration probes vector modes generated by scalar perturbations of the same wavelength.  It was shown in~\cite{Lu:2007cj} that vectors modes are only generated by 2 scalar modes when the scalar modes have differing wavelength, otherwise no angular momentum is generated by the interaction (this is not the case for the tensors). 

For the tensors the kernel is
\bea
\mathcal{K}_{\mathrm{T}}^{(2)} (\bk_{1},\bk_{2},\bk_{3}) &=&
3\Omega_m\cH^4 \bigg\{\bigg[
-3\Omega_m(1+\mathcal{Q})-2f^2\mathcal{Q}-(3\Omega_m-2f^2)\frac{\cH'}{\cH^2}
+4\frac{ff'}{\cH}
\bigg]\mathcal{G}(\eta,k_3)\nonumber\\&&
+\frac{3\Omega_m+2f^2}{\cH}\mathcal{G}'(\eta,k_3)
\bigg\}
\frac{{e^{ij}(\bm k_3)k_{1i}k_{2j}\,\,n^ln^me_{lm}(\bm k_3)}
+{\bar e^{ij}(\bm k_3)k_{1i}k_{2j}\,\,n^ln^m\bar e_{lm}(\bm k_3)}}{k_1^2k_2^2k_3^2}\,,
\eea
where $\mathcal{G}(\eta,k)$ arises from solving the tensor wave equation and is given in the Appendix~\eqref{alskdjbdbhv}. Here, $e_{ij}$ and $\bar e_{ij}$ are orthogonal polarization tensors. For both kernels the products of the polarizations with $k$ vectors is complicated and needs to be done in a coordinate system. This results in different expressions for each permutation~-- see the Appendices for full details.

\subsection{Bispectrum}

We can now give an expression for the local vector and tensor contributions to the galaxy bispectrum in terms of the respective kernels presented above. We drop the time dependence of $\Delta_{g}$ as we only calculate  the equal-time bispectrum. At second order, the bispectrum is defined after symmetrizing over external momenta:
\begin{align}
\frac{1}{2}\big\langle \Delta_{g}^{(1)}(\bk_{1})\Delta_{g}^{(1)}(\bk_{2})\Delta^{(2)}_{g}(\bk_{3}) \big\rangle+ \mathrm{2\;cyc.\;perm.}=(2\pi)^3B_{g}(\bk_{1},\bk_{2},\bk_{3})\delta^{D}(\bk_{1}+\bk_{2}+\bk_{3})\;.
\label{eq:15}
\end{align}
Using Wick's theorem, we can write the bispectrum in terms of the kernels as,
\begin{equation}
B_{g}(\bm{k}_{1},\bm{k}_{2},\bm{k}_{3}) = \mathcal{K}^{(1)}(\bm{k}_{1})\mathcal{K}^{(1)}(\bm{k}_{2})\mathcal{K}^{(2)}(\bm{k}_{1}, \bm{k}_{2}, \bm{k}_{3}) P(k_{1}) P(k_{2}) + \mathrm{2\;cyc.\;perm.} \;, \label{Bisp}
\end{equation}
where $P$ is the linear dark matter power spectrum. The first-order kernel has only scalars (S) and is given by \citep{Umeh:2016nuh}:
\begin{equation}
\mathcal{K}^{(1)}(\bk_{1})= b_{1}+f\mu_{1}^{2}+\mathrm{i}\mu_{1}\frac{\gamma_{1}}{k_{1}} + \frac{\gamma_{2}}{k_{1}^{2}}\;.
\end{equation}
The Newtonian part is given by the bias plus redshift-space distortion term, $\mathcal{K}^{(1)}_{\mathrm{N}}(\bk_{1})=b_{1}+f\mu_{1}^{2}$.
The $\gamma$-coefficients arise from GR contributions and are given by \citep{Jolicoeur:2017nyt}:
\begin{eqnarray}
{\gamma_1\over \mathcal{H}} &=&  { f}\bigg[b_{e}  - 2\mathcal{Q} - \frac{2(1-\mathcal{Q})}{\chi\mathcal{H}}- \frac{\mathcal{H}'}{\mathcal{H}^{2}} \bigg]\;,
\\ \label{gamma1}
\frac{\gamma_{2}}{\mathcal{H}^{2}} &=& f(3-b_{e}) +\frac{3}{2}\Omega_{m} \left[2+b_{e}-f-4\mathcal{Q}-2 \frac{\left(1- \mathcal{Q} \right)}{\chi\mathcal{H}}-\frac{\mathcal{H}'}{\mathcal{H}^2}\right]\;. \label{gamma2}
\end{eqnarray}
whilst the second-order kernel has scalars, vectors (V) and tensors (T) i.e., 
\begin{equation}
\mathcal{K}^{(2)} = \mathcal{K}^{(2)}_{\mathrm{S}} + \mathcal{K}^{(2)}_{\mathrm{V}} + \mathcal{K}^{(2)}_{\mathrm{T}}\;. \label{Kernel2}
\end{equation}
The scalar contribution has both Newtonian and scalar GR projection effects \citep{Umeh:2016nuh}. For {Gaussian} initial conditions the galaxy bispectrum is induced by second order perturbations only. The computation of the galaxy bispectrum from scalars including first- and second-order GR contributions has already been done in \citep{Umeh:2016nuh, Jolicoeur:2017nyt, Jolicoeur:2017eyi}. Here, we focus on the vector and tensor contributions which are given as follows:
\begin{eqnarray}
B_{g\mathrm{V}}(\bk_{1},\bk_{2},\bk_{3}) &=& \mathcal{K}^{(1)}(\k_{1})\mathcal{K}^{(1)}(\k_{2})\mathcal{K}^{(2)}_{\mathrm{V}} (\bk_{1},\bk_{2},\bk_{3})P(k_{2})P(k_{1}) + \mathrm{2\;cyc.\;perm.}\;, \label{eq:16} \\
B_{g\mathrm{T}}(\bk_{1},\bk_{2},\bk_{3}) &=& \mathcal{K}^{(1)}(\k_{1})\mathcal{K}^{(1)}(\k_{2}) \mathcal{K}^{(2)}_{\mathrm{T}} (\bk_{1},\bk_{2},\bk_{3})P(k_{2})P(k_{1})+\mathrm{2\;cyc.\;perm.}\;.
\label{eq:16b}
\end{eqnarray}
Similarly, the full scalar contribution to the bispectrum is defined by:
\begin{equation}
\label{eq:ScalBispec}
B_{g\mathrm{S}}(\bk_{1},\bk_{2},\bk_{3}) = \mathcal{K}^{(1)}(\k_{1})\mathcal{K}^{(1)}(\k_{2}) \mathcal{K}^{(2)}_{\mathrm{S}} (\bk_{1},\bk_{2},\bk_{3})P(k_{2})P(k_{1}) + \mathrm{2\;cyc.\;perm.}\, . 
\end{equation}
In each of these expressions we keep the full $\mathcal{O}(1)$ GR kernel. For reference we also plot the pure Newtonian bispectrum. 

As the bispectrum is a complex function of several variables it has many properties which cannot be captured on a single graph. In the Newtonian case it is a purely real function. To compare with this we compute the monopole of the bispectrum,
\begin{equation}
\label{eq:Monopole}
B_{g}^{0}(k_{1}, k_{2}, k_{3}) = 
\frac{1}{\sqrt{4\pi}}\int_{0}^{2\pi}\ud \phi \int_{-1}^{+1} \ud \mu_{1}\,B_{g}(k_{1}, k_{2}, k_{3}, \mu_{1}, \phi, \phi_{n})\,. 
\end{equation}
This function gives the mean bispectrum averaged over the sky and some angles in Fourier space, and the sign of it tells us about whether over-densities appear more or less clustered relative to voids than in a Gaussian random field.   
The monopole is a purely real function by symmetry of the bispectrum, so in computing the monopole we automatically ignore new relativistic contributions which arise only in the imaginary part of the bispectrum, and signal odd parity symmetries in apparent clustering. In order to see the importance of these we also compute the monopole of the absolute value of the bispectrum,
\begin{equation}
\label{eq:MonopoleAbs}
B_{g}^{\text{abs},0}(k_{1}, k_{2}, k_{3}) = 
\frac{1}{\sqrt{4\pi}}\int_{0}^{2\pi}\ud \phi \int_{-1}^{+1} \ud \mu_{1}\,|B_{g}(k_{1}, k_{2}, k_{3}, \mu_{1}, \phi, \phi_{n})|\,. 
\end{equation}
In the Newtonian case this has contributions from even multipoles of the bispectrum, as well as summing up the total bispectrum amplitude. Because this includes contributions from all multipoles, this function  gives a better overall picture of the amplitude of the bispectrum than the monopole itself. In the case of the relativistic bispectrum, this will indicate where new relativistic contributions will be important. To isolate the effects of vector and tensor modes on the galaxy bispectrum, and to compare them with the Newtonian contributions, we construct the analogues of the monopole (\ref{eq:Monopole}) and of the monopole of the absolute value (\ref{eq:MonopoleAbs}) of the whole bispectrum by replacing the bispectrum $B_{g}(k_{1}, k_{2}, k_{3}, \mu_{1}, \phi, \phi_{n})$ in Eq.~(\ref{eq:Monopole}) and Eq.~(\ref{eq:MonopoleAbs}) by the expressions~(\ref{eq:16})-(\ref{eq:ScalBispec}) or by the Newtonian bispectrum.
We consider  the monopoles at $z=1.0$, and we use the following numerical values for the other parameters for illustration purposes \citep{Jolicoeur:2017eyi}:
\begin{itemize}
\item $b_{1}=\sqrt{1+z}$, $b_{2}=-0.1 \sqrt{1+z}$ and $b_{e}=\mathcal{Q} = 0$.
\item We include tidal bias with $b_s = -4(b_1-1)/7$.
\item $\Omega_{m0} = 1-\Omega_{\Lambda0} = 0.308$ and $h = 0.678$.
\item For the shape of the galaxy bispectrum, we consider the moderately squeezed triangle ($k_{1} = k_{2} = k$ and $k_{3} \approx k/16$).
\end{itemize}

\begin{figure}[ht]
\centering
\includegraphics[width=0.7\textwidth] {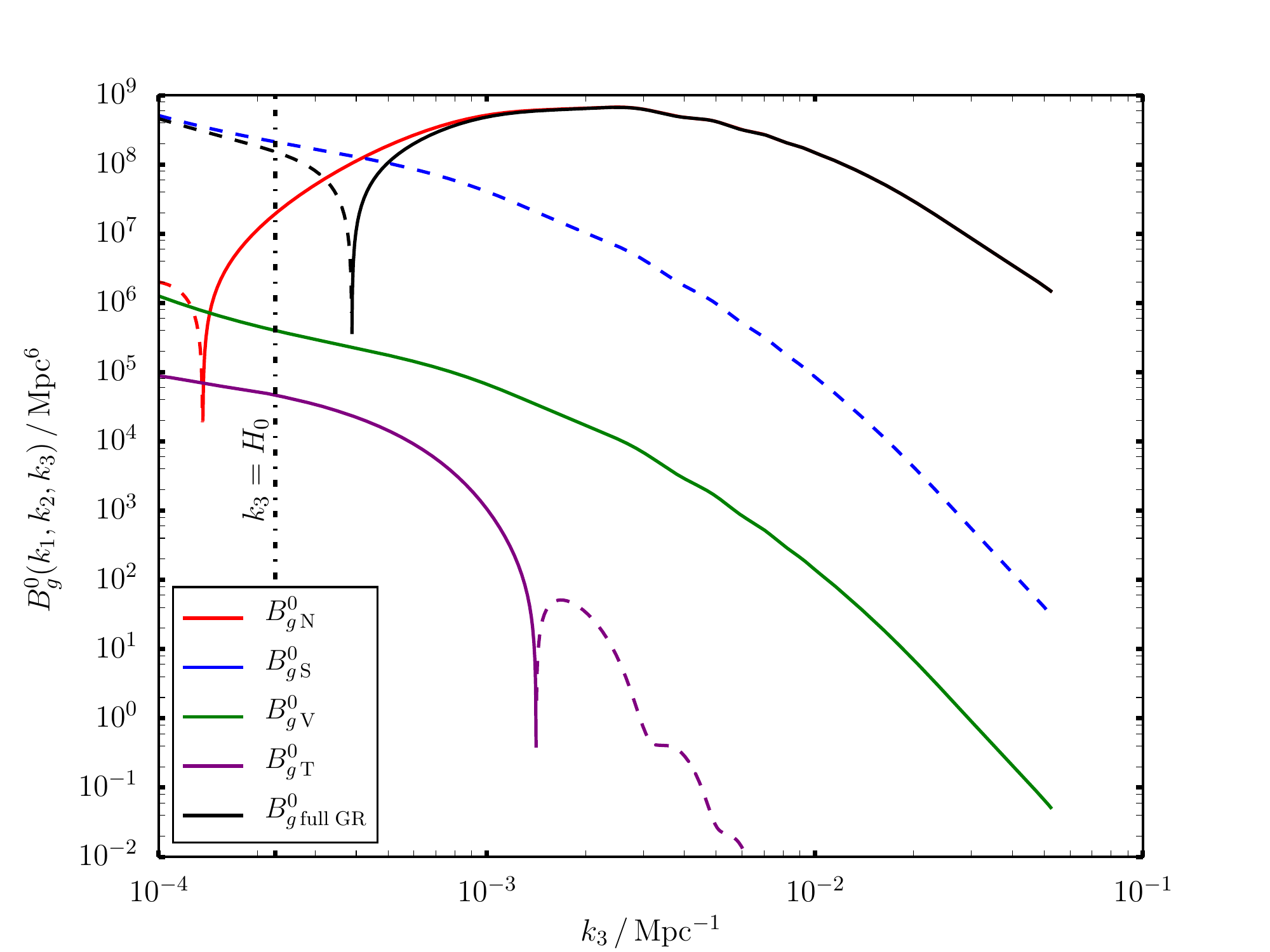}
\caption{The monopole at $z=1.0$ for the moderately squeezed configuration.  Dashed lines indicate negative.} \label{fig1}
\end{figure}
\begin{figure}[ht]
\centering
\includegraphics[width=0.7\textwidth] {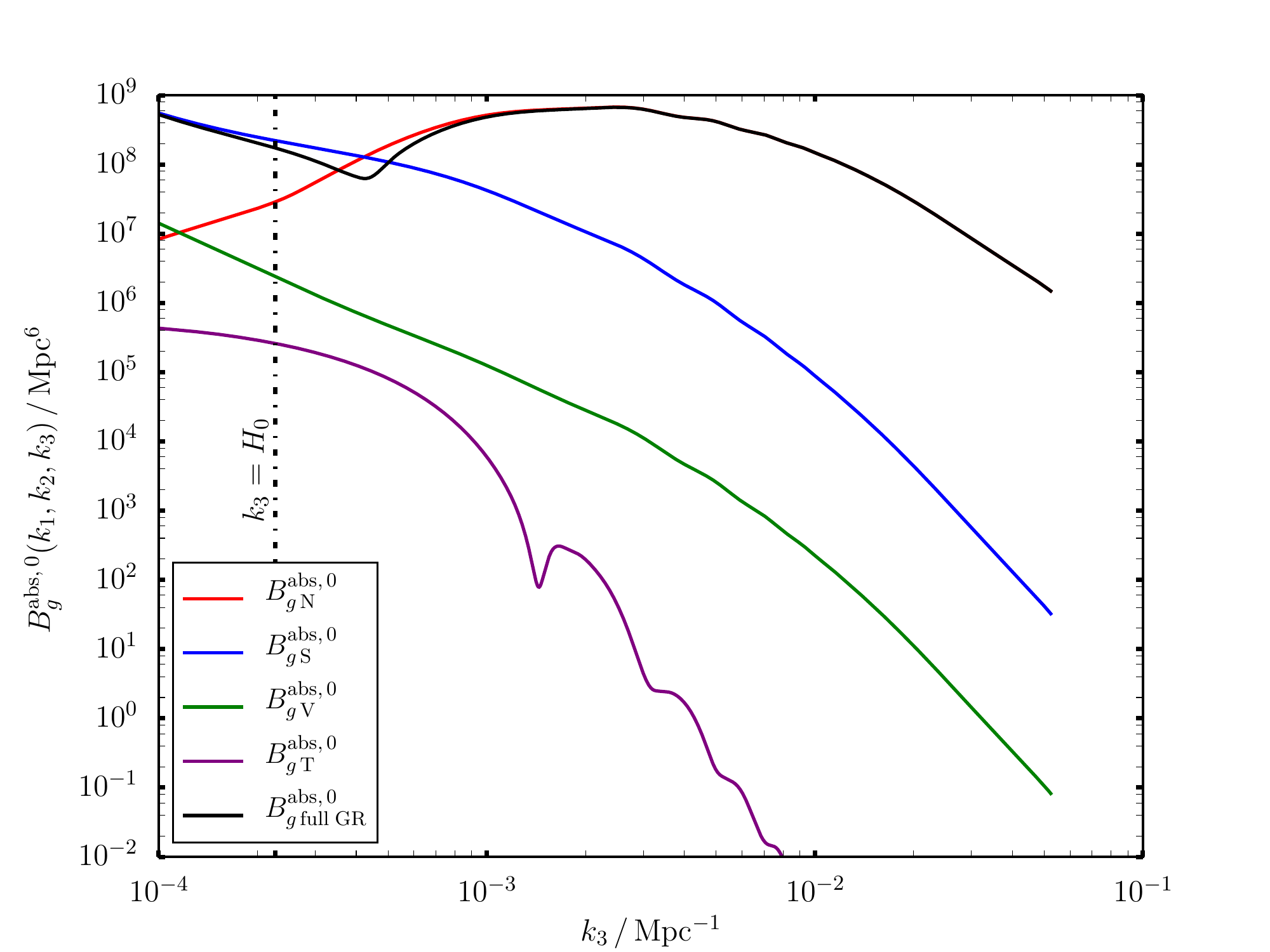} 
\caption{The monopole of the absolute value of the bispectrum at $z=1.0$ for the moderately squeezed configuration.  } \label{fig2}
\end{figure}
In Figs.~\ref{fig1} and \ref{fig2} we show the vector and tensor contributions to the bispectrum, for both monopoles. For reference we also show the scalar contribution, and the pure Newtonian bispectrum. Finally we show in black the total bispectrum with all contributions, given by
\begin{equation}
B_{g\,\mathrm{full\;GR}}(\bm{k}_{1},\bm{k}_{2},\bm{k}_{3}) = \mathcal{K}^{(1)}(\bm{k}_{1})\mathcal{K}^{(1)}(\bm{k}_{2})\mathcal{K}^{(2)}(\bm{k}_{1},\bm{k}_{2},\bm{k}_{3})P(k_{1})P(k_{2})\;+\;\text{2 cycl. perm.}\;.
\end{equation}

First consider Fig.~\ref{fig1}. These show that the vectors and tensors are small on all scales as expected, only growing towards the Hubble scale, because their leading behaviour scales as $(\mathcal{H}/k)^2$. We see that while the vector contribution to the pure monopole is significantly larger than the tensors, it remains about 3 orders of magnitude below the scalar GR contribution. When the long mode is outside the Hubble radius (and unobservable), the vector part does become larger than the Newtonian contribution, but this is only because of the effect of tidal bias making the Newtonian monopole negative for very large modes. The scalar GR effect on the monopole is negative on all scales, making the full bispectrum negative on sub-Hubble scales~-- in contrast the Newtonian result is positive in this region. Note that the tensors oscillate on small scales, picking up the baryon acoustic oscillations. 

Turning now to Fig~\ref{fig2}, we see the effect of the vectors and tensors are relatively enhanced compared to the scalar and Newtonian (though still sub-dominant). In fact the vectors are enhanced by over an order of magnitude because their dominant contribution is in the anti-symmetric imaginary part (which has no monopole). 

\section{conclusions}

We have considered for the first time the relativistic contributions of vectors and tensors on the observed galaxy bispectrum. We have derived our results using a geometry suitable for any source of vectors and tensors, and then specialised to the case of vectors and tensors generated by scalar coupling at first order which simplifies the geometry of the kernels (azimuthal angles only appear as $\phi_n-\phi$ meaning the polarizations do not need integrated out for such a source). We present our results for a moderately squeezed configuration, finding that the local vectors and tensors (as defined in the Poisson gauge) are small and can safely be neglected in future bispectrum studies. We have not included the integrated contributions from these sources which may alter this conclusion.  In contrast, the scalars cannot be neglected.

\acknowledgements 

CC was supported by STFC Consolidated Grant ST/P000592/1. JL is supported by the National Research Foundation (South Africa). OU is supported by STFC grant ST/N000668/1. RM is supported by STFC Consolidated Grant ST/N000668/1, and by  the South African Radio Astronomy Observatory (SARAO) and the National Research Foundation (Grant No. 75415).

\appendix

\section{Derivation of the kernels for local vectors and tensors}
\label{AppA}

\subsection{Vector bispectrum}
Let us consider an observer receiving a photon along the  direction $\bm{n}$ in its rest frame.
Neglecting integrated terms and terms at the observation point, we obtain the contribution of second order local vector modes to the observed galaxy number density as:
{
\begin{align}
\Delta_{g\mathrm{V}}^{(2)}(z, \bm{x})= \bigg[b_{e}-2\mathcal{Q} - \frac{2(1-\mathcal{Q})}{\chi\cH} - \frac{\cH'}{\cH^{2}}\bigg]\hat{v}^{(2)}_{\|} - \frac{1}{\cH}\partial_{\|}\hat{v}^{(2)}_{\|}\;,
\label{eq:App1}
\end{align}
}
where {$\hat{v}_{\|}^{(2)}=n^{i}\hat{v}_{i}^{(2)}$} is the longitudinal component of the vector perturbations. All the expressions are implicitly evaluated at the observed redshift. 
Note that we  have ${\hat{v}_{i}^{(2)}}=-\omega_{i}^{(2)}$ in a $\Lambda$CDM Universe and hence
{
\begin{align}
\Delta_{g\mathrm{V}}^{(2)}(z, \bm{x})=\left[-b_{e}+2\mathcal{Q} + \frac{2(1-\mathcal{Q})}{\chi\cH} + \frac{\cH'}{\cH^{2}}+\frac{1}{\mathcal{H}}\partial_{\|} \right] \omega^{(2)}_{\|}(\bm{x})\;.
\label{eq:App2}
\end{align}
}
where $\omega_{\|}^{(2)}=n^{i}\omega_{i}^{(2)}$. The Fourier transform of a pure vector can be represented in terms of two independent polarization vectors as \citep{Lu:2008ju}:
\begin{equation}
\omega_{i}^{(2)}(\bm{x})=\int \frac{\ud^{3}k}{(2\pi)^3} \big[\omega (\bk) e_{i}(\bk)+\bar{\omega}(\bk) \bar{e}_{i}(\bk)\big]\mathrm{e}^{\mathrm{i}\bk\cdot\bm{x}}\;.
\label{eq:App3}
\end{equation}
with, 
\begin{equation}
\omega (\bk) = \int \ud^{3}x\,\omega_{i}(\bm{x})e^{i}(\bm{k})\mathrm{e}^{-\mathrm{i}\bm{k}\cdot \bm{x}}, \qquad \bar{\omega} (\bk) = \int \ud^{3}x\,\omega_{i}(\bm{x})\bar{e}^{i}(\bm{k})\mathrm{e}^{-\mathrm{i}\bm{k}\cdot \bm{x}} \label{eq:wk}
\end{equation}
Taking \eqref{eq:App3} we write
\begin{equation}
\omega_{\|}^{(2)}(\bm{x}) = n^{i}\omega_{i}^{(2)}(\bm{x})=\int \frac{\ud^{3}k}{(2\pi)^3} \big[\omega (\bk) n^{i}e_{i}(\bk)+\bar{\omega}(\bk) n^{i}\bar{e}_{i}(\bk)\big]\mathrm{e}^{\mathrm{i}\bk\cdot\bm{x}}\;, \label{eq:omegai1}
\end{equation}
{
The Fourier transform of \eqref{eq:wi} gives, 
\begin{equation}
\omega_{i}(\bm{k}_{3}) = -\mathrm{i}\frac{6\Omega_{m}\cH^{3}f}{k_{3}^{2}}\int\ud(\bm{k}_{1}, \bm{k}_{2}, \bm{k}_{3})\,\bigg[\frac{k_{1i}}{k_{1}^{2}} + \frac{k_{2i}}{k_{2}^{2}}\bigg]\;. \label{A6}
\end{equation}
Then,
\begin{equation}
\omega(\bm{k}_{3}) = e^{i}(\bm{k}_{3})\omega_{i}(\bm{k}_{3}) = -\mathrm{i}\frac{6\Omega_{m}\cH^{3}f}{k_{3}^{2}}\int\ud(\bm{k}_{1}, \bm{k}_{2}, \bm{k}_{3})\,\bigg[\frac{\bm{k}_{1} \cdot \bm{e}(\bm{k}_{3})}{k_{1}^{2}} + \frac{\bm{k}_{2} \cdot \bm{e}(\bm{k}_{3})}{k_{2}^{2}}\bigg]\;. \label{A7}
\end{equation} 
}
The other parity $\bar{\omega}(\bm{k})$, has similar Fourier solutions with $\bm{e}(\bm{k})$ being replaced by $\bm{\bar{e}}(\bm{k})$. We also have
\begin{equation}
\Delta_{g\mathrm{V}}^{(2)}\left(\bm{x}\right) = \int \frac{\ud^{3}k_{3}}{(2\pi)^3}\,\Delta_{g\mathrm{V}}^{(2)}\left(\bk_{3}\right) \mathrm{e}^{\mathrm{i}\bk_{3}\cdot\bm{x}}
\end{equation}
where  
\begin{eqnarray}
\Delta_{g\mathrm{V}}^{(2)}\left(\bm{k}_{3}\right) &=& \int\ud(\bm{k}_{1}, \bm{k}_{2}, \bm{k}_{3})\,\mathcal{K}_{\mathrm{V}}^{(2)} (\bm{k}_{1},\bm{k}_{2},\bm{k}_{3})\;, \label{eq:kernelV1} 
\end{eqnarray}
and $\mathcal{K}_{\mathrm{V}}^{(2)}$ is the kernel for the vector perturbations. 
These can be written as
\begin{eqnarray}
\mathcal{K}^{(2)}_{\mathrm{V}}(\bm{k}_{1}, \bm{k}_{2}, \bm{k}_{3} ) &=& 12\Omega_{m}\cH^{2}f\frac{\mu_{3}}{k_{3}}\mathcal{V}_{123} 
-12 \mathrm{i}\Omega_{m}\cH^{3}f\bigg[-b_{e}+2\mathcal{Q}+\frac{2(1-\mathcal{Q})}{\chi\cH}+\frac{\cH'}{\cH^{2}}\bigg]
\frac{\mathcal{V}_{123}}{k_{3}^{2}} \nonumber\\
&=& 
12\Omega_{m}\cH^{2}\left[f{\mu_{3}}
-\mathrm{i}\frac{\gamma_1}{k_3}\right]
\frac{\mathcal{V}_{123}}{k_{3}} \;, \label{e2} 
\end{eqnarray} 
where we define
\be
\mathcal{V}_{123}=\left[
\bm n\cdot \bm e(\bk_3)\left(\frac{\bk_1\cdot\bm e(\bk_3)}{k_1^2}+\frac{\bk_2\cdot\bm e(\bk_3)}{k_2^2}
\right)
+\bm n\cdot \bar{\bm e}(\bk_3)\left(\frac{\bk_1\cdot\bar{\bm e}(\bk_3)}{k_1^2}+\frac{\bk_2\cdot\bar{\bm e}(\bk_3)}{k_2^2}
\right)
\right]
\ee
and $\gamma_1$ is defined below. 
At this stage we need to introduce coordinates in Fourier space~-- we specify how we do this in Appendix~\ref{appB}. This breaks the algebraic cyclic symmetry of the kernel, so we work out the exact form of the kernels, giving,
\begin{eqnarray}
\mathcal{V}_{123} &=& \left(\frac{k_{2}^{2} - k_{1}^{2}}{k_{3}^{2}k_{1}}\right)
\sqrt{1-\mu_{12}^2}\left[
\mu_1\sqrt{1-\mu_{12}^2}-\left(\frac{k_1}{k_2}+\mu_{12}\right)\sqrt{1-\mu_1^2}\cos(\phi_n-\phi)
\right]\;, \label{v123}\\
\mathcal{V}_{231} &=& \left(\frac{k_{3}^{2}-k_{2}^{2}}{k_{3}^{2}k_{2}}\right)\sqrt{1-\mu_{1}^{2}}\sqrt{1-\mu_{12}^{2}}{\cos{(\phi_n-\phi)}}\;, \label{v231} \\ 
\mathcal{V}_{312} &=& \left(\frac{k_{3}^{2} - k_{1}^{2}}{k_{3}^{2}k_{1}}\right) 
\sqrt{1-\mu_{12}^2}\left[
\mu_1\sqrt{1-\mu_{12}^2}-\mu_{12}\sqrt{1-\mu_1^2}\cos(\phi_n-\phi)
\right]
\label{v312}
\end{eqnarray}

Finally, for Gaussian initial conditions, we obtain the vector contribution to the bispectrum:
\begin{align}
\small
B_{g\mathrm{V}}(\bk_{1},\bk_{2},\bk_{3})=\mathcal{K}^{(1)}(\k_{1})\mathcal{K}^{(1)}(\k_{2}) \mathcal{K}^{(2)}_{\mathrm{V}} (\bk_{1},\bk_{2},\bk_{3})P(k_{1})P(k_{2})+\mathrm{2\;cyc.\;perm.}\;.
\label{eq:App9}
\end{align}

\subsection{Tensor bispectrum}
We expand in Fourier space using the polarization tensors $e_{ij}$ and $\bar{e}_{ij}$ which are  defined as \citep{Hwang:2017oxa, Ananda:2006af},
\begin{equation}
e_{ij}(\bm{k}) = \frac{1}{\sqrt{2}}\big[e_{i}(\bm{k}) e_{j}(\bm{k}) - \bar{e}_{i}(\bm{k})\bar{e}_{j}(\bm{k})\big]\;, \qquad \bar{e}_{ij}(\bm{k}) = \frac{1}{\sqrt{2}}\big[e_{i}(\bm{k})\bar{e}_{j}(\bm{k}) + \bar{e}_{i}(\bm{k})e_{j}(\bm{k})\big]\;. \label{polTen}
\end{equation}
Then,
\begin{equation}
h_{ij}^{(2)}(\bm{x})=\int \frac{\ud^{3}k}{(2\pi)^3} \big[h (\bk) e_{ij}(\bk)+\bar{h}(\bk) \bar{e}_{ij}(\bk)\big]\mathrm{e}^{\mathrm{i}\bk\cdot\bm{x}}\;.
\end{equation}
with, 
\begin{equation}
h (\bk) = \int \ud^{3}x\,h_{ij}(\bm{x})e^{ij}(\bm{k})\mathrm{e}^{-\mathrm{i}\bm{k}\cdot \bm{x}}, \qquad \bar{h} (\bk) = \int \ud^{3}x\,h_{ij}(\bm{x})\bar{e}^{ij}(\bm{k})\mathrm{e}^{-\mathrm{i}\bm{k}\cdot \bm{x}} \label{eq:wk}
\end{equation}
In Fourier space we then have 
\begin{equation}
h_\parallel (\bk)=h(\bk)e_{ij}(\bk)n^{i}n^{j}+\bar{h}(\bk)\bar{e}_{ij}(\bk))n^{i}n^{j}\;, \label{ten:1}
\end{equation}
which obeys
\be
h''(\eta,\bk)-2\cH h'(\eta,\bk)+k^2 h(\eta,\bk)=S(\eta,\bk)\,,
\ee
$S(\bk)$ is the source term in the evolution equation of $h(\bm{k})$ which is obtained from the transverse, trace-free part of the Einstein field equations~\eqref{sdjncsdjkcn}.
The galaxy density fluctuations become
\begin{align}
\Delta^{(2)}_{g\mathrm{T}}(\bk)=-\frac{1}{2}\left({1-\mathcal{Q}}+\frac{1}{\mathcal{H}}\frac{\partial}{\partial \eta}\right)\Big[h(\bk){e_{ij}}(\bk)n^{i}n^{j}+\bar{h}(\bk)\bar{e}_{ij}(\bk)n^{i}n^{j}\Big]\;. \label{ten:2}
\end{align}
For simplicity we give the solution for the tensors using the approximation that we are in the matter era. 
In general, the solution for the second order tensor is written as an integral over the Green's function written in terms of the spherical Bessel functions, \citep{Baumann:2007zm},
\begin{align}
h(\eta,\bk)=\frac{1}{a}\int^\eta\ud\tilde\eta\, \frac{\eta\tilde\eta}{k^2}\,\left[j_1(k\eta)y_1(k\tilde\eta)-j_1(k\tilde\eta)y_1(k\eta)\right]a(\tilde\eta)S(\tilde\eta,\bk)\;. \label{h3}
\end{align}
However, since $S(\eta,\bk)$ is nearly constant in time in the matter era at high redshift, we can approximate this as~\cite{Hwang:2017oxa}
\begin{align}
h(\eta,\bk)\approx\frac{{G}(\eta,k)}{k^2}S(\bk)\;, \label{h3}
\end{align}
where ${G}(\eta,k)$ is given by , 
\begin{align}\label{alskdjbdbhv}
{G}(\eta,k)=1+\frac{3\big[k\eta\cos{(k\eta)}-\sin{(k\eta)}]}{(k\eta)^3}\;.
\end{align}
 
The source appearing in the Fourier transformed tensor equation becomes,
\be
S(\bm k) = -6\Omega_m\cH^4\left(3\Omega_m+2f^2\right) \int \ud(\bm{k}_{1}, \bm{k}_{2}, \bm{k})
e^{ij}(\bm k)\frac{k_{1i}k_{2j}}{k_1^2k_2^2}\,.
\ee
In this approximation, we have ignored the weak time dependence (from $g$) in the source when integrating over the Green's function to obtain \eqref{h3}. At high redshift this should be a very good approximation. The other parity is given by the same expression with the other polarisation tensor. 

As with the vectors, in Fourier space we write the contribution to the galaxy density contrast as an integral over a kernel, 
\be
\Delta^{(2)}_{g\mathrm{T}}(\bk_{3}) = \int \ud(\bm{k}_{1}, \bm{k}_{2}, \bm{k}_{3}) \, \mathcal{K}_{\mathrm{T}}^{(2)} (\bk_{1},\bk_{2},\bk_{3})\;, \label{eq:13}
\ee
and we find:
\bea
\mathcal{K}_{\mathrm{T}}^{(2)} (\bk_{1},\bk_{2},\bk_{3}) &=&
3\Omega_m\cH^4 \bigg\{\bigg[
-3\Omega_m(1+\mathcal{Q})-2f^2\mathcal{Q}-(3\Omega_m-2f^2)\frac{\cH'}{\cH^2}
+4\frac{ff'}{\cH}
\bigg]\mathcal{G}(\eta,k_3)\nonumber\\&&
+\frac{3\Omega_m+2f^2}{\cH}\mathcal{G}'(\eta,k_3)
\bigg\}
\frac{{\cal T}_{123}}{k_3^2}
\eea
where we defined
\be
{\cal T}_{123} = \frac{1}{k_1^2k_2^2}\bigg[{e^{ij}(\bm k_3)k_{1i}k_{2j}\,\,n^ln^me_{lm}(\bm k_3)}
+{\bar e^{ij}(\bm k_3)k_{1i}k_{2j}\,\,n^ln^m\bar e_{lm}(\bm k_3)}\bigg]\,.
\ee
As with the vectors, the 123 symmetry is broken on choosing a coordinate system in Fourier space, so we give the permutations explicitly as
{
\begin{eqnarray}
\mathcal{T}_{123} &=& \frac{1}{4k_{3}^{4}}\big(1-\mu_{12}^{2}\big) \Big[k_{3}^{2}-k_{1}^{2}+\big(k_{1}^{2}-2k_{2}^{2}-k_{3}^{2}\big)\mu_{1}^{2} + \big(2k_{1}k_{2}\mu_{1}^{2}-2k_{1}k_{2}\big)\mu_{12} + \big(3k_{2}^{2}\mu_{1}^{2}-k_{2}^{2}\big)\mu_{12}^{2} \nonumber \\
&&{} \qquad \qquad \qquad \; +4k_{2}\big(k_{1}+k_{2}\mu_{12}\big)\mu_{1}\sqrt{1-\mu_{1}^{2}}\sqrt{1-\mu_{12}^{2}}\cos{(\phi_{n}-\phi)} \nonumber \\
&&{} \qquad \qquad \qquad \;- \big(1-\mu_{1}^{2}\big)\big(k_{1}^{2}+2k_{1}k_{2}\mu_{12}+k_{2}^{2}\mu_{12}^{2}+k_{3}^{2}\big)\cos{2(\phi_n - \phi)}\Big]\;, \label{A31} \\ 
\mathcal{T}_{231} &=& -\frac{1}{2k_{3}^{2}}\big(1-\mu_{1}^{2}\big)\big(1-\mu_{12}^{2}\big)\cos{2(\phi_{n}-\phi)}\;, \label{A32} \\
\mathcal{T}_{312} &=& \frac{1}{4k_{3}^{2}}\big(1-\mu_{12}^{2}\big)\Big[\big(1-\mu_{12}^{2}\big)\big(1-3\mu_{1}^{2}\big) + 4\mu_{1}\mu_{12}\sqrt{1-\mu_{1}^{2}}\sqrt{1-\mu_{12}^{2}}\cos{(\phi_n - \phi)} \nonumber \\
&&{} \qquad \qquad \qquad \;-\big(1-\mu_{1}^{2}\big)\big(1+\mu_{12}^{2}\big)\cos{2(\phi_n - \phi)}\Big]\;. \label{A33}
\end{eqnarray}
}
The tensor contribution to the bispectrum then reads:
\begin{align}
B_{g\mathrm{T}}(\bk_{1},\bk_{2},\bk_{3})= \mathcal{K}^{(1)}(\k_{1})\mathcal{K}^{(1)}(\k_{2})\mathcal{K}^{(2)}_{\mathrm{T}} (\bk_{1},\bk_{2},\bk_{3}) P(k_{1})P(k_{2})+\mathrm{2\;cyc.\;perm.}\;. \label{A34}
\end{align}

\section{Geometry}\label{appB}

Here we define the coordinates in Fourier space in which our Fourier space polarisation vectors are decomposed, which we show in Fig.~\ref{fig3}.

First we consider the relations among the  $\bk$-vectors, the direction vector $\bm n$, and the polarisation vectors and tensors we use:
\begin{itemize}
\item The cosines of the angles $\theta_{i}$ and $\theta_{ij}$ are defined as  
\begin{equation}
\cos{\theta_{i}} = \mu_{i} = \hat{\bm{k}}_{i} \cdot \hat{\bm{n}}\;, \quad \cos{\theta_{ij}}=\mu_{ij} = \hat{\bm{k}}_{i} \cdot \hat{\bm{k}}_{j}\;. \label{e7}
\end{equation}
\item The wave vectors form a triangle,
$\bm{k}_{1}+\bm{k}_{2}+\bm{k}_{3}={0}$, and by taking the dot product with $\bm{n}$ we have
\begin{equation}
\mu_{1}k_{1} + \mu_{2}k_{2} + \mu_{3}k_{3} = 0\;.\label{e10}
\end{equation}
\item 
Now using $k_{3}^{2} = k_{1}^{2} + k_{2}^{2} + 2k_{1}k_{2}\mu_{12}$
and if we define $r=k_{2}/k_1 $, we have $k_{3} = k_{1}\sqrt{1 + r^{2} + 2r\mu_{12}}$. Using the cosine rule on the other sides of the triangle we have 
\begin{equation}
\mu_{23} = \frac{-(r+\mu_{12})}{\sqrt{1 + r^{2} + 2r\mu_{12}}}\;, \qquad \mu_{31} = \frac{-(1+r\mu_{12})}{\sqrt{1 + r^{2} + 2r\mu_{12}}}\;. \label{c13}
\end{equation}
\item Taking dot-products of the polarisation vectors with the triangle of $\bk$-vectors, we have
\begin{eqnarray}
\bm{k}_{1} \cdot \bm{e}(\bm{k}_{3}) = - \bm{k}_{2} \cdot \bm{e}(\bm{k}_{3})\;, \qquad \bm{k}_{2}\cdot \bm{e}(\bm{k}_{1}) = -\bm{k}_{3}\cdot \bm{e}(\bm{k}_{1})\;, \qquad \bm{k}_{3} \cdot \bm{e}(\bm{k}_{2}) = - \bm{k}_{1} \cdot \bm{e}(\bm{k}_{2})\;, \label{polaeq1} \\ 
\bm{k}_{1} \cdot \bm{\bar{e}}(\bm{k}_{3}) = - \bm{k}_{2} \cdot \bm{\bar{e}}(\bm{k}_{3})\;, \qquad \bm{k}_{2}\cdot \bm{\bar{e}}(\bm{k}_{1}) = -\bm{k}_{3}\cdot \bm{\bar{e}}(\bm{k}_{1}), \qquad \bm{k}_{3} \cdot \bm{\bar{e}}(\bm{k}_{2}) = - \bm{k}_{1} \cdot \bm{\bar{e}}(\bm{k}_{2})\;. \label{polaeq2}
\end{eqnarray}
\item The polarization tensors $e_{ij}$ and $\bar{e}_{ij}$ are defined as \citep{Hwang:2017oxa, Ananda:2006af},
\begin{equation}
e_{ij}(\bm{k}) = \frac{1}{\sqrt{2}}\big[e_{i}(\bm{k}) e_{j}(\bm{k}) - \bar{e}_{i}(\bm{k})\bar{e}_{j}(\bm{k})\big]\;, \qquad \bar{e}_{ij}(\bm{k}) = \frac{1}{\sqrt{2}}\big[e_{i}(\bm{k})\bar{e}_{j}(\bm{k}) + \bar{e}_{i}(\bm{k})e_{j}(\bm{k})\big]\;. \label{polTen}
\end{equation}

\end{itemize}

Now we set up our coordinate system. 
\begin{figure}[! ht]
\centering
 \hspace{-1.5cm}\includegraphics[width=0.6\textwidth]{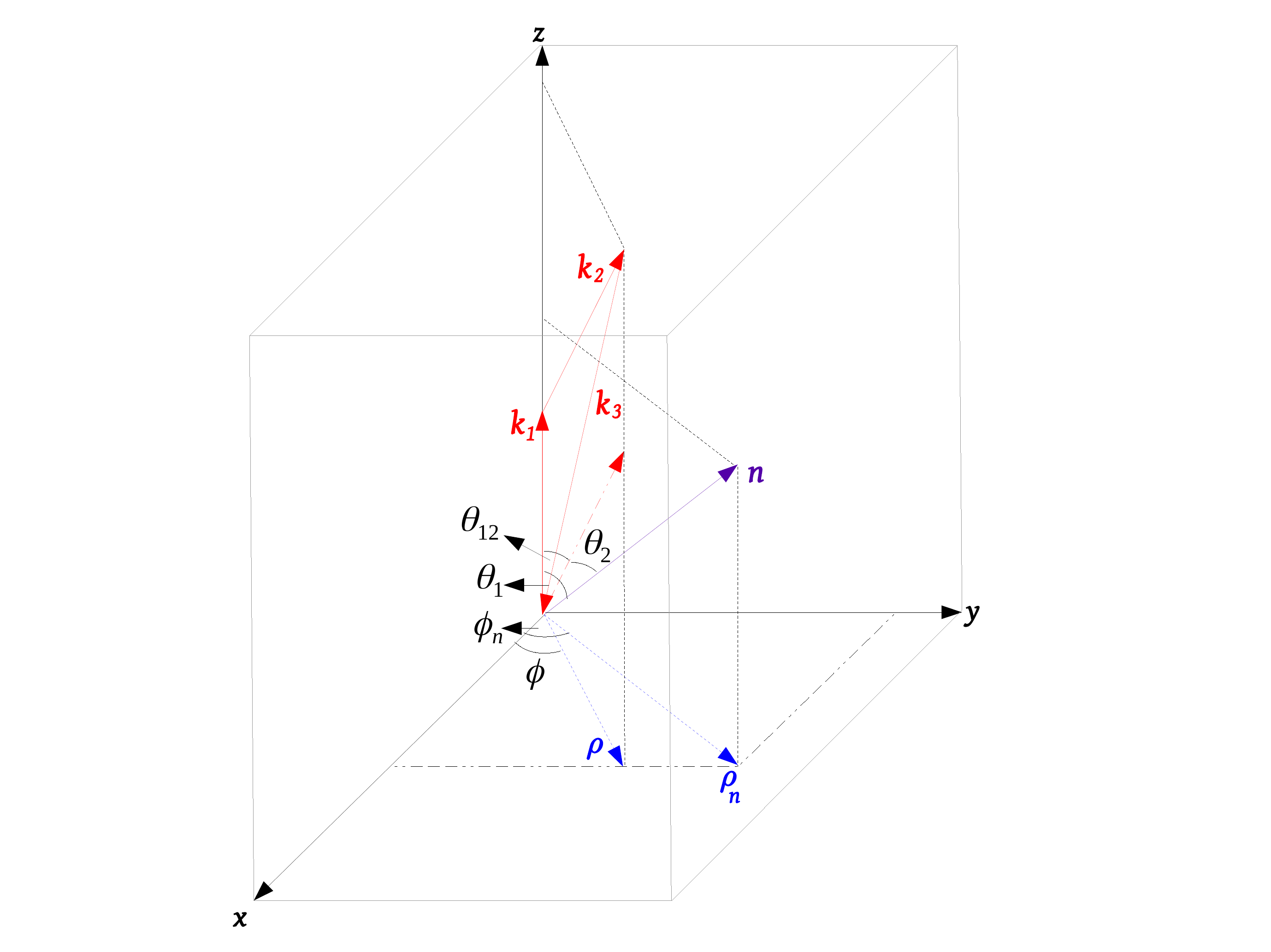} \hspace{-2.5cm}
\includegraphics[width=0.6\textwidth]{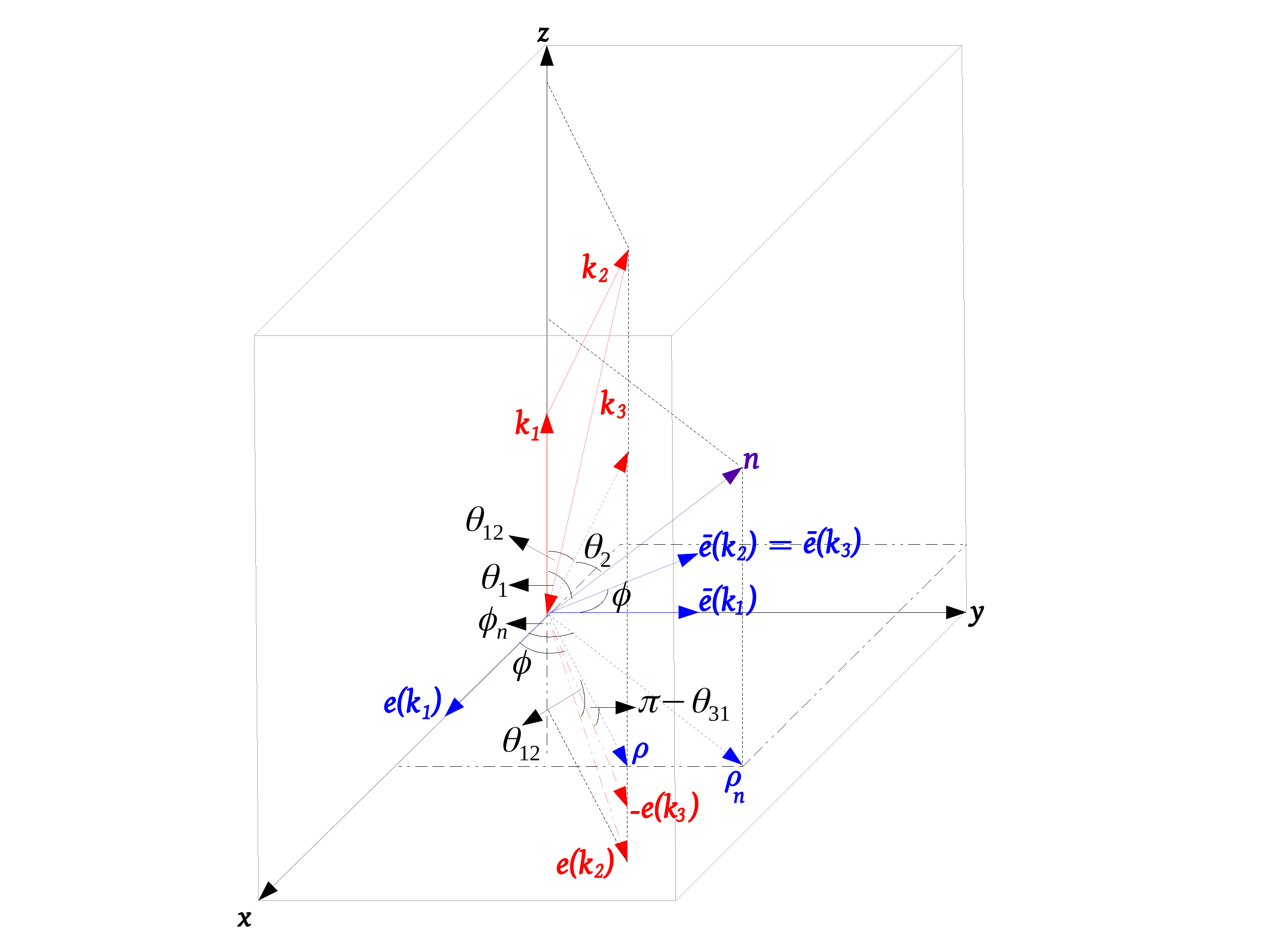}
\caption{The geometry of the galaxy bispectrum without (\emph{left}) and with (\emph{right}) polarizations.}\label{fig3}
\end{figure}
\begin{itemize}
\item We choose our coordinates to coincide with the $\bk_1$ system, with $\bk_1$ lying along the $z$-axis, and $\bm{e}(\bk_1)$ along $x$. Then
\be
\bm{e}(\bk_1)=(1,0,0),~~~\bar{\bm{e}}(\bk_1)=(0,1,0),~~~\hat\bk_1=(0,0,1)
\ee 
\item We define the azimuthal angle $\bm n$ makes with the $x$-axis as $\phi_n$, which implies, using $\theta_1$ as its polar angle,
\be
\bm n = (\sin\theta_1\cos\phi_n,\sin\theta_1\sin\phi_n,\cos\theta_1)
\ee
\item We define the azimuthal angle $\bk_2$ makes with the $x$-axis as $\phi$, implying
\be
\bm k_2 = k_2(\sin\theta_{12}\cos\phi,\sin\theta_{12}\sin\phi,\cos\theta_{12})
\ee
 This implies $\mu_2$ is given by	
\begin{equation}
\mu_{2} = \mu_{1}\mu_{12} + \sqrt{1-\mu_{1}^{2}}\sqrt{1-\mu_{12}^{2}}\cos{(\phi_{n}-\phi)}\;. \label{e8}
\end{equation}

\item For $\bk_3$ we write
\be
\bm k_3 = k_3(-\sin\theta_{31}\cos\phi,-\sin\theta_{31}\sin\phi,\cos\theta_{31})
\ee
and using the fact that the wavevectors form a triangle, we find $\sin\theta_{31}=\frac{k_2}{k_3}\sin\theta_{12}$, giving
\be
\bm k_3 = -k_2\bigg(\sin\theta_{12}\cos\phi,\sin\theta_{12}\sin\phi,\frac{k_1}{k_2}+\cos\theta_{12}\bigg)\,,
\ee

\item { To define $(\bm e,\bar{\bm e})(\bk_2)$ we choose the triplet $[\bm{k}_{2},\bm{e}(\bm{k}_{2}),\bm{\bar{e}}(\bm{k}_{2})]$ to coincide with $[\bm{k}_{1},\bm{e}(\bm{k}_{1}),\bm{\bar{e}}(\bm{k}_{1})]$ when $\theta_{12} = \phi = 0$ (refer to Fig. \ref{fig1}).} This implies
\begin{eqnarray}
\bm{e}(\bm{k}_{2}) &=& (\cos{\theta_{12}}\cos{\phi}, \cos{\theta_{12}}\sin{\phi}, -\sin{\theta_{12}})\;, \label{B12}\\
\bm{\bar{e}}(\bm{k}_{2}) &=& (-\sin{\phi}, \cos{\phi},0)\;, \label{B13} 
\end{eqnarray}
 where the direction is chosen so that $(\bm e,\bar{\bm e})(\bk_2)$ has the same handedness as $(\bm e,\bar{\bm e})(\bk_1)$ (with respect to their $\bk$ vectors).
 
 \item We specify $(\bm e,\bar{\bm e})(\bk_3)$ in the same way, giving
 \bea
 \bm{e}(\bm{k}_{3}) &=& -\frac{1}{k_{3}}\Big(\big[k_{1}+k_{2}\cos{\theta_{12}}\big]\cos{\phi}, \big[k_{1}+k_{2}\cos{\theta_{12}}\big]\sin{\phi}, -k_{2}\sin{\theta_{12}}\Big)\;, \label{B15} \\
 \bm{\bar{e}}(\bm{k}_{3}) &=& (-\sin{\phi},\cos{\phi}, 0)\;. \label{B16}
 \eea
 This implies
 \be
 \bk_2\cdot{\bm e}(\bk_3) = -\frac{k_1k_2}{k_3}\sqrt{1-\mu_{12}^2}\,.
 \ee
 \item Gathering these results together, for the dot products between the mode vectors and polarization vectors we have,
\begin{eqnarray}
\bm{k}_{1} \cdot \bm{e}(\bm{k}_{2}) &=& -k_{1}\sqrt{1-\mu_{12}^{2}}\;, \label{B18} \\
\bm{k}_{1} \cdot \bm{\bar{e}}(\bm{k}_{2}) &=& 0\;, \label{B19} \\
\bm{k}_{2} \cdot \bm{e}(\bm{k}_{1}) &=& k_{2}\sqrt{1-\mu_{12}^{2}}\cos{\phi}\;, \label{B20} \\
\bm{k}_{2} \cdot \bm{\bar{e}}(\bm{k}_{1}) &=& k_{2}\sqrt{1-\mu_{12}^{2}}\sin{\phi}\;, \label{B21} \\
\bm{k}_{2} \cdot \bm{e}(\bm{k}_{3}) &=& -\frac{k_{1}k_{2}}{k_{3}}\sqrt{1-\mu_{12}^{2}}\;, \label{B22} \\
\bm{k}_{2} \cdot \bm{\bar{e}}(\bm{k}_{3}) &=& 0\;. \label{B23}
\end{eqnarray}

\item For the tensors we require
\bea
e^{ij}(\bm k_3)k_{1i}k_{2j} &=& -\frac{k_1^2k_2^2}{\sqrt{2}k_3^2}(1-\mu_{12}^2)\;, \label{B24} \\
\bar e^{ij}(\bm k_3)k_{1i}k_{2j} &=& 0\;, \label{B25} \\
e^{ij}(\bm k_2)k_{1i}k_{3j} &=& -\frac{1}{\sqrt{2}}k_1^2(1-\mu_{12}^2)\; , \label{B26} \\
\bar e^{ij}(\bm k_2)k_{1i}k_{3j} &=& 0\;, \label{B27}\\
e^{ij}(\bm k_1)k_{2i}k_{3j} &=&  -\frac{1}{\sqrt{2}}k_2^2(1-\mu_{12}^2)\cos 2\phi  \;, \label{B28}\\
\bar e^{ij}(\bm k_1)k_{2i}k_{3j} &=&-\frac{1}{\sqrt{2}}k_2^2(1-\mu_{12}^2)\sin 2\phi  \;. \label{B29} \\
\eea
 
 \item We then have relations with the observation direction
\begin{eqnarray}
\bm{n} \cdot \bm{e}(\bm{k}_{1}) &=& \sqrt{1-\mu_{1}^{2}}\cos{\phi_{n}}\;, \label{B31} \\
\bm{n} \cdot \bm{\bar{e}}(\bm{k}_{1}) &=& \sqrt{1-\mu_{1}^{2}}\sin{\phi_{n}}\;, \label{B32} \\
\bm{n} \cdot \bm{e}(\bm{k}_{2}) &=& \mu_{12}\sqrt{1-\mu_{1}^{2}}\cos{(\phi_{n}-\phi)}-\mu_{1}\sqrt{1-\mu_{12}^{2}}\;, \label{B33} \\
\bm{n} \cdot \bm{\bar{e}}(\bm{k}_{2}) &=& \sqrt{1-\mu_{1}^{2}}\sin{(\phi_{n}-\phi)}\;, \label{B34} \\
\bm{n} \cdot \bm{e}(\bm{k}_{3}) &=& \frac{k_{2}}{k_{3}}\bigg[-\sqrt{1-\mu_{1}^{2}}\bigg(\frac{k_{1}}{k_{2}}+\mu_{12}\bigg)\cos{(\phi_{n}-\phi)} + \mu_{1}\sqrt{1-\mu_{12}^{2}}\bigg]\;, \label{B35} \\
\bm{n} \cdot \bm{\bar{e}}(\bm{k}_{3}) &=& \sqrt{1-\mu_{1}^{2}}\sin{(\phi_{n}-\phi)}\;. \label{B36} 
\end{eqnarray}

\item For the polarisation tensors contracted along the line of sight, we have, 
\begin{eqnarray}
\bm{e}_{ij}(\bm{k}_{1})n^{i}n^{j} &=& \frac{1}{\sqrt{2}}(1-\mu_{1}^{2})\cos{2\phi_{n}}\;, \label{B37} \\
\bm{\bar{e}}_{ij}(\bm{k}_{1})n^{i}n^{j} &=& \frac{1}{\sqrt{2}}(1-\mu_{1}^{2})\sin{2\phi_{n}}\;, \label{B38} \\
\bm{e}_{ij}(\bm{k}_{2})n^{i}n^{j} &=& \frac{1}{\sqrt{2}}\bigg[\big(1-\mu_{12}^{2}\big)\sin^{2}(\phi_{n}-\phi) - \mu_{12}^{2}\big(1-\mu_{1}^{2}\big)\cos^{2}(\phi_{n}-\phi) + 2\mu_{1}\mu_{12}\sqrt{1-\mu_{1}^{2}}\sqrt{1-\mu_{12}^{2}}\cos{(\phi_{n}-\phi)} \nonumber \\
&& \qquad \quad- \mu_{1}^{2}\big(1-\mu_{12}^{2}\big)\bigg] \;, \label{B39} \\
\bm{\bar{e}}_{ij}(\bm{k}_{2})n^{i}n^{j} &=& \frac{1}{\sqrt{2}}\bigg[\mu_{12}\big(1-\mu_{1}^{2}\big)\sin{2(\phi_{n}-\phi)}-2\mu_{1}\sqrt{1-\mu_{1}^{2}}\sqrt{1-\mu_{12}^{2}}\sin{(\phi_{n}-\phi)}\bigg]\;, \label{B40} \\
\bm{e}_{ij}(\bm{k}_{3})n^{i}n^{j} &=& \frac{1}{\sqrt{2}}\bigg\{\big(1-\mu_{1}^{2}\big)\sin^{2}{(\phi_{n}-\phi)} - \frac{k_{2}^{2}}{k_{3}^{2}}\bigg[\bigg(\frac{k_{1}}{k_{2}}+\mu_{12}\bigg)^{2}\big(1-\mu_{1}^{2}\big)\cos^{2}{(\phi_{n}-\phi)}+ \mu_{1}^{2}\big(1-\mu_{12}^{2}\big)\bigg] \nonumber \\
&&{} \qquad \qquad \qquad \qquad \qquad \qquad \qquad \quad -2\bigg(\frac{k_{1}}{k_{2}}+\mu_{12}\bigg)\mu_{1}\sqrt{1-\mu_{1}^{2}}\sqrt{1-\mu_{12}^{2}}\cos{(\phi_{n}-\phi)} \bigg]\bigg\}\;, \label{B41} \\
\bm{\bar{e}}_{ij}(\bm{k}_{3})n^{i}n^{j} &=& \frac{1}{\sqrt{2}}\frac{k_{2}}{k_{3}}\bigg[-\big(1-\mu_{1}^{2}\big)\bigg(\frac{k_{1}}{k_{2}}+\mu_{12}\bigg)\sin{2(\phi_{n}-\phi)} + 2\mu_{1}\sqrt{1-\mu_{1}^{2}}\sqrt{1-\mu_{12}^{2}}\sin{(\phi_{n}-\phi)}\bigg]\;. \label{B42}
\end{eqnarray}

\end{itemize}

\bibliographystyle{JHEP}  
\bibliography{reference_library}

\providecommand{\href}[2]{#2}\begingroup\raggedright\begin{thebibliography}{10}

\bibitem{Sefusatti:2006pa}
E.~Sefusatti, M.~Crocce, S.~Pueblas, and R.~Scoccimarro, {\it {Cosmology and
  the Bispectrum}},  {\em Phys. Rev.} {\bf D74} (2006) 023522,
  [\href{http://arxiv.org/abs/astro-ph/0604505}{{\tt astro-ph/0604505}}].

\bibitem{Sefusatti:2007ih}
E.~Sefusatti and E.~Komatsu, {\it {The bispectrum of galaxies from
  high-redshift galaxy surveys: Primordial non-Gaussianity and non-linear
  galaxy bias}},  {\em Phys. Rev.} {\bf D76} (2007) 083004,
  [\href{http://arxiv.org/abs/0705.0343}{{\tt arXiv:0705.0343}}].

\bibitem{Komatsu:2010hc}
E.~Komatsu, {\it {Hunting for Primordial Non-Gaussianity in the Cosmic
  Microwave Background}},  {\em Class. Quant. Grav.} {\bf 27} (2010) 124010,
  [\href{http://arxiv.org/abs/1003.6097}{{\tt arXiv:1003.6097}}].

\bibitem{Liguori:2010hx}
M.~Liguori, E.~Sefusatti, J.~R. Fergusson, and E.~P.~S. Shellard, {\it
  {Primordial non-Gaussianity and Bispectrum Measurements in the Cosmic
  Microwave Background and Large-Scale Structure}},  {\em Adv. Astron.} {\bf
  2010} (2010) 980523, [\href{http://arxiv.org/abs/1001.4707}{{\tt
  arXiv:1001.4707}}].

\bibitem{Yadav:2010fz}
A.~P.~S. Yadav and B.~D. Wandelt, {\it {Primordial Non-Gaussianity in the
  Cosmic Microwave Background}},  {\em Adv. Astron.} {\bf 2010} (2010) 565248,
  [\href{http://arxiv.org/abs/1006.0275}{{\tt arXiv:1006.0275}}].

\bibitem{Baldauf:2010vn}
T.~Baldauf, U.~Seljak, and L.~Senatore, {\it {Primordial non-Gaussianity in the
  Bispectrum of the Halo Density Field}},  {\em JCAP} {\bf 1104} (2011) 006,
  [\href{http://arxiv.org/abs/1011.1513}{{\tt arXiv:1011.1513}}].

\bibitem{Yokoyama:2013mta}
S.~Yokoyama, T.~Matsubara, and A.~Taruya, {\it {Halo/galaxy bispectrum with
  primordial non-Gaussianity from integrated perturbation theory}},  {\em Phys.
  Rev.} {\bf D89} (2014), no.~4 043524,
  [\href{http://arxiv.org/abs/1310.4925}{{\tt arXiv:1310.4925}}].

\bibitem{Camera:2014bwa}
S.~Camera, M.~G. Santos, and R.~Maartens, {\it {Probing primordial
  non-Gaussianity with SKA galaxy redshift surveys: a fully relativistic
  analysis}},  {\em Mon. Not. Roy. Astron. Soc.} {\bf 448} (2015), no.~2
  1035--1043, [\href{http://arxiv.org/abs/1409.8286}{{\tt arXiv:1409.8286}}].

\bibitem{Abramo:2017xnp}
L.~R. Abramo and D.~Bertacca, {\it {Disentangling the effects of Doppler
  velocity and primordial non-Gaussianity in galaxy power spectra}},  {\em
  Phys. Rev.} {\bf D96} (2017), no.~12 123535,
  [\href{http://arxiv.org/abs/1706.01834}{{\tt arXiv:1706.01834}}].

\bibitem{Karagiannis:2018jdt}
D.~Karagiannis, A.~Lazanu, M.~Liguori, A.~Raccanelli, N.~Bartolo, and L.~Verde,
  {\it {Constraining Primordial non-Gaussianity with Bispectrum and Power
  Spectum from Upcoming Optical and Radio Surveys}},
  \href{http://arxiv.org/abs/1801.09280}{{\tt arXiv:1801.09280}}.

\bibitem{Creminelli:2013nua}
P.~Creminelli, J.~Gleyzes, L.~Hui, M.~Simonović, and F.~Vernizzi, {\it
  {Single-Field Consistency Relations of Large Scale Structure. Part III: Test
  of the Equivalence Principle}},  {\em JCAP} {\bf 1406} (2014) 009,
  [\href{http://arxiv.org/abs/1312.6074}{{\tt arXiv:1312.6074}}].

\bibitem{Hashimoto:2017klo}
I.~Hashimoto, Y.~Rasera, and A.~Taruya, {\it {Precision cosmology with
  redshift-space bispectrum: a perturbation theory based model at one-loop
  order}},  {\em Phys. Rev.} {\bf D96} (2017), no.~4 043526,
  [\href{http://arxiv.org/abs/1705.02574}{{\tt arXiv:1705.02574}}].

\bibitem{Tellarini:2015faa}
M.~Tellarini, A.~J. Ross, G.~Tasinato, and D.~Wands, {\it {Non-local bias in
  the halo bispectrum with primordial non-Gaussianity}},  {\em JCAP} {\bf 1507}
  (2015), no.~07 004, [\href{http://arxiv.org/abs/1504.00324}{{\tt
  arXiv:1504.00324}}].

\bibitem{Tellarini:2016sgp}
M.~Tellarini, A.~J. Ross, G.~Tasinato, and D.~Wands, {\it {Galaxy bispectrum,
  primordial non-Gaussianity and redshift space distortions}},  {\em JCAP} {\bf
  1606} (2016), no.~06 014, [\href{http://arxiv.org/abs/1603.06814}{{\tt
  arXiv:1603.06814}}].

\bibitem{DiDio:2016gpd}
E.~Di~Dio, H.~Perrier, R.~Durrer, G.~Marozzi, A.~Moradinezhad~Dizgah,
  J.~Noreña, and A.~Riotto, {\it {Non-Gaussianities due to Relativistic
  Corrections to the Observed Galaxy Bispectrum}},  {\em JCAP} {\bf 1703}
  (2017), no.~03 006, [\href{http://arxiv.org/abs/1611.03720}{{\tt
  arXiv:1611.03720}}].

\bibitem{Challinor:2011bk}
A.~Challinor and A.~Lewis, {\it {The linear power spectrum of observed source
  number counts}},  {\em Phys. Rev.} {\bf D84} (2011) 043516,
  [\href{http://arxiv.org/abs/1105.5292}{{\tt arXiv:1105.5292}}].

\bibitem{Bonvin:2011bg}
C.~Bonvin and R.~Durrer, {\it {What galaxy surveys really measure}},  {\em
  Phys. Rev.} {\bf D84} (2011) 063505,
  [\href{http://arxiv.org/abs/1105.5280}{{\tt arXiv:1105.5280}}].

\bibitem{Bonvin:2014owa}
C.~Bonvin, {\it {Isolating relativistic effects in large-scale structure}},
  {\em Class. Quant. Grav.} {\bf 31} (2014), no.~23 234002,
  [\href{http://arxiv.org/abs/1409.2224}{{\tt arXiv:1409.2224}}].

\bibitem{DiDio:2015bua}
E.~Di~Dio, R.~Durrer, G.~Marozzi, and F.~Montanari, {\it {The bispectrum of
  relativistic galaxy number counts}},  {\em JCAP} {\bf 1601} (2016) 016,
  [\href{http://arxiv.org/abs/1510.04202}{{\tt arXiv:1510.04202}}].

\bibitem{Kehagias:2015tda}
A.~Kehagias, A.~Moradinezhad~Dizgah, J.~Noreña, H.~Perrier, and A.~Riotto,
  {\it {A Consistency Relation for the Observed Galaxy Bispectrum and the Local
  non-Gaussianity from Relativistic Corrections}},  {\em JCAP} {\bf 1508}
  (2015), no.~08 018, [\href{http://arxiv.org/abs/1503.04467}{{\tt
  arXiv:1503.04467}}].

\bibitem{Bertacca:2017dzm}
D.~Bertacca, A.~Raccanelli, N.~Bartolo, M.~Liguori, S.~Matarrese, and L.~Verde,
  {\it {Relativistic wide-angle galaxy bispectrum on the light-cone}},  {\em
  Phys. Rev.} {\bf D97} (2018), no.~2 023531,
  [\href{http://arxiv.org/abs/1705.09306}{{\tt arXiv:1705.09306}}].

\bibitem{Koyama:2018ttg}
K.~Koyama, O.~Umeh, R.~Maartens, and D.~Bertacca, {\it {The observed galaxy
  bispectrum from single-field inflation in the squeezed limit}},  {\em JCAP}
  {\bf 1807} (2018), no.~07 050, [\href{http://arxiv.org/abs/1805.09189}{{\tt
  arXiv:1805.09189}}].

\bibitem{Umeh:2016nuh}
O.~Umeh, S.~Jolicoeur, R.~Maartens, and C.~Clarkson, {\it {A general
  relativistic signature in the galaxy bispectrum: the local effects of
  observing on the lightcone}},  {\em JCAP} {\bf 1703} (2017) 003,
  [\href{http://arxiv.org/abs/1610.03351}{{\tt arXiv:1610.03351}}].

\bibitem{Jolicoeur:2017nyt}
S.~Jolicoeur, O.~Umeh, R.~Maartens, and C.~Clarkson, {\it {Imprints of local
  lightcone projection effects on the galaxy bispectrum. II}},  {\em JCAP} {\bf
  1709} (2017) 040, [\href{http://arxiv.org/abs/1703.09630}{{\tt
  arXiv:1703.09630}}].

\bibitem{Jolicoeur:2017eyi}
S.~Jolicoeur, O.~Umeh, R.~Maartens, and C.~Clarkson, {\it {Imprints of local
  lightcone projection effects on the galaxy bispectrum. Part III. Relativistic
  corrections from nonlinear dynamical evolution on large-scales}},  {\em JCAP}
  {\bf 1803} (2018), no.~03 036, [\href{http://arxiv.org/abs/1711.01812}{{\tt
  arXiv:1711.01812}}].

\bibitem{Tomita:2005et}
K.~Tomita, {\it {Relativistic second-order perturbations of nonzero-lambda flat
  cosmological models and CMB anisotropies}},  {\em Phys. Rev.} {\bf D71}
  (2005) 083504, [\href{http://arxiv.org/abs/astro-ph/0501663}{{\tt
  astro-ph/0501663}}].

\bibitem{Mollerach:2003nq}
S.~Mollerach, D.~Harari, and S.~Matarrese, {\it {CMB polarization from
  secondary vector and tensor modes}},  {\em Phys. Rev.} {\bf D69} (2004)
  063002, [\href{http://arxiv.org/abs/astro-ph/0310711}{{\tt
  astro-ph/0310711}}].

\bibitem{Ananda:2006af}
K.~N. Ananda, C.~Clarkson, and D.~Wands, {\it {The Cosmological gravitational
  wave background from primordial density perturbations}},  {\em Phys. Rev.}
  {\bf D75} (2007) 123518, [\href{http://arxiv.org/abs/gr-qc/0612013}{{\tt
  gr-qc/0612013}}].

\bibitem{Lu:2007cj}
T.~H.-C. Lu, K.~Ananda, and C.~Clarkson, {\it {Vector modes generated by
  primordial density fluctuations}},  {\em Phys. Rev.} {\bf D77} (2008) 043523,
  [\href{http://arxiv.org/abs/0709.1619}{{\tt arXiv:0709.1619}}].

\bibitem{Baumann:2007zm}
D.~Baumann, P.~J. Steinhardt, K.~Takahashi, and K.~Ichiki, {\it {Gravitational
  Wave Spectrum Induced by Primordial Scalar Perturbations}},  {\em Phys. Rev.}
  {\bf D76} (2007) 084019, [\href{http://arxiv.org/abs/hep-th/0703290}{{\tt
  hep-th/0703290}}].

\bibitem{Lu:2008ju}
T.~H.-C. Lu, K.~Ananda, C.~Clarkson, and R.~Maartens, {\it {The cosmological
  background of vector modes}},  {\em JCAP} {\bf 0902} (2009) 023,
  [\href{http://arxiv.org/abs/0812.1349}{{\tt arXiv:0812.1349}}].

\bibitem{Hwang:2017oxa}
J.-C. Hwang, D.~Jeong, and H.~Noh, {\it {Gauge dependence of gravitational
  waves generated from scalar perturbations}},  {\em Astrophys. J.} {\bf 842}
  (2017), no.~1 46, [\href{http://arxiv.org/abs/1704.03500}{{\tt
  arXiv:1704.03500}}].

\bibitem{Bruni:2013mua}
M.~Bruni, D.~B. Thomas, and D.~Wands, {\it {Computing General Relativistic
  effects from Newtonian N-body simulations: Frame dragging in the
  post-Friedmann approach}},  {\em Phys. Rev.} {\bf D89} (2014), no.~4 044010,
  [\href{http://arxiv.org/abs/1306.1562}{{\tt arXiv:1306.1562}}].

\bibitem{Adamek:2015eda}
J.~Adamek, D.~Daverio, R.~Durrer, and M.~Kunz, {\it {General relativity and
  cosmic structure formation}},  {\em Nature Phys.} {\bf 12} (2016) 346--349,
  [\href{http://arxiv.org/abs/1509.01699}{{\tt arXiv:1509.01699}}].

\bibitem{Andrianomena:2014sya}
S.~Andrianomena, C.~Clarkson, P.~Patel, O.~Umeh, and J.-P. Uzan, {\it
  {Non-linear relativistic contributions to the cosmological weak-lensing
  convergence}},  {\em JCAP} {\bf 1406} (2014) 023,
  [\href{http://arxiv.org/abs/1402.4350}{{\tt arXiv:1402.4350}}].

\bibitem{Saga:2015apa}
S.~Saga, D.~Yamauchi, and K.~Ichiki, {\it {Weak lensing induced by second-order
  vector mode}},  {\em Phys. Rev.} {\bf D92} (2015), no.~6 063533,
  [\href{http://arxiv.org/abs/1505.02774}{{\tt arXiv:1505.02774}}].

\bibitem{Durrer:2016jzq}
R.~Durrer and V.~Tansella, {\it {Vector perturbations of galaxy number
  counts}},  {\em JCAP} {\bf 1607} (2016), no.~07 037,
  [\href{http://arxiv.org/abs/1605.05974}{{\tt arXiv:1605.05974}}].

\bibitem{Bonvin:2017req}
C.~Bonvin, R.~Durrer, N.~Khosravi, M.~Kunz, and I.~Sawicki, {\it
  {Redshift-space distortions from vector perturbations}},  {\em JCAP} {\bf
  1802} (2018), no.~02 028, [\href{http://arxiv.org/abs/1712.00052}{{\tt
  arXiv:1712.00052}}].

\bibitem{Tansella:2018hdm}
V.~Tansella, C.~Bonvin, G.~Cusin, R.~Durrer, M.~Kunz, and I.~Sawicki, {\it
  {Redshift-space distortions from vector perturbations II: Anisotropic
  signal}},  \href{http://arxiv.org/abs/1807.00731}{{\tt arXiv:1807.00731}}.

\bibitem{Bojowald:2007hv}
M.~Bojowald and G.~M. Hossain, {\it {Cosmological vector modes and quantum
  gravity effects}},  {\em Class. Quant. Grav.} {\bf 24} (2007) 4801--4816,
  [\href{http://arxiv.org/abs/0709.0872}{{\tt arXiv:0709.0872}}].

\bibitem{Dai:2013kra}
L.~Dai, D.~Jeong, and M.~Kamionkowski, {\it {Anisotropic imprint of
  long-wavelength tensor perturbations on cosmic structure}},  {\em Phys. Rev.}
  {\bf D88} (2013), no.~4 043507, [\href{http://arxiv.org/abs/1306.3985}{{\tt
  arXiv:1306.3985}}].

\bibitem{Matarrese:1993zf}
S.~Matarrese, O.~Pantano, and D.~Saez, {\it {General relativistic dynamics of
  irrotational dust: Cosmological implications}},  {\em Phys. Rev. Lett.} {\bf
  72} (1994) 320--323, [\href{http://arxiv.org/abs/astro-ph/9310036}{{\tt
  astro-ph/9310036}}].

\bibitem{Noh:2004bc}
H.~Noh and J.-c. Hwang, {\it {Second-order perturbations of the Friedmann world
  model}},  {\em Phys. Rev.} {\bf D69} (2004) 104011.

\bibitem{Malik:2008im}
K.~A. Malik and D.~Wands, {\it {Cosmological perturbations}},  {\em Phys.
  Rept.} {\bf 475} (2009) 1--51, [\href{http://arxiv.org/abs/0809.4944}{{\tt
  arXiv:0809.4944}}].

\bibitem{Hwang:2012aa}
J.-c. Hwang and H.~Noh, {\it {Fully nonlinear and exact perturbations of the
  Friedmann world model}},  {\em Mon. Not. Roy. Astron. Soc.} {\bf 433} (2013)
  3472, [\href{http://arxiv.org/abs/1207.0264}{{\tt arXiv:1207.0264}}].

\bibitem{Rampf:2013dxa}
C.~Rampf, {\it {Frame dragging and Eulerian frames in General Relativity}},
  {\em Phys. Rev.} {\bf D89} (2014) 063509,
  [\href{http://arxiv.org/abs/1307.1725}{{\tt arXiv:1307.1725}}].

\bibitem{Osano:2015pea}
B.~Osano, {\it {Second Order perturbation Theory: A covariant approach
  involving a barotropic equation of state}},  {\em Class. Quant. Grav.} {\bf
  34} (2017), no.~12 125004, [\href{http://arxiv.org/abs/1504.01495}{{\tt
  arXiv:1504.01495}}].

\bibitem{Rampf:2014mga}
C.~Rampf and A.~Wiegand, {\it {Relativistic Lagrangian displacement field and
  tensor perturbations}},  {\em Phys. Rev.} {\bf D90} (2014) 123503,
  [\href{http://arxiv.org/abs/1409.2688}{{\tt arXiv:1409.2688}}].

\bibitem{Bertacca:2014dra}
D.~Bertacca, R.~Maartens, and C.~Clarkson, {\it {Observed galaxy number counts
  on the lightcone up to second order: I. Main result}},  {\em JCAP} {\bf 1409}
  (2014), no.~09 037, [\href{http://arxiv.org/abs/1405.4403}{{\tt
  arXiv:1405.4403}}].

\bibitem{Bertacca:2014wga}
D.~Bertacca, R.~Maartens, and C.~Clarkson, {\it {Observed galaxy number counts
  on the lightcone up to second order: II. Derivation}},  {\em JCAP} {\bf 1411}
  (2014), no.~11 013, [\href{http://arxiv.org/abs/1406.0319}{{\tt
  arXiv:1406.0319}}].

\bibitem{Bertacca:2014hwa}
D.~Bertacca, {\it {Observed galaxy number counts on the light cone up to second
  order: III. Magnification bias}},  {\em Class. Quant. Grav.} {\bf 32} (2015),
  no.~19 195011, [\href{http://arxiv.org/abs/1409.2024}{{\tt
  arXiv:1409.2024}}].

\bibitem{Nielsen:2016ldx}
J.~T. Nielsen and R.~Durrer, {\it {Higher order relativistic galaxy number
  counts: dominating terms}},  {\em JCAP} {\bf 1703} (2017), no.~03 010,
  [\href{http://arxiv.org/abs/1606.02113}{{\tt arXiv:1606.02113}}].

\end{thebibliography}\endgroup

\end{document}